\documentclass[10pt,twocolumn,letterpaper]{article}

\usepackage{iccv}
\usepackage{times}
\usepackage{epsfig}
\usepackage{graphicx}
\usepackage{amsmath}
\usepackage{amssymb}

\usepackage[accsupp]{axessibility}  

\usepackage{float}
\usepackage{fixltx2e}
\usepackage[font=small,skip=0pt]{caption}
\usepackage{epstopdf}
\usepackage{booktabs}
\def\be{\begin{equation}}\def\ee{\end{equation}}
\def\ba{\begin{array}}\def\ea{\end{array}}
\def\bfg{\begin{figure}}\def\efg{\end{figure}}
\usepackage[breaklinks=true,bookmarks=false]{hyperref}

\iccvfinalcopy 

\def\iccvPaperID{31} 

\ificcvfinal\fi

\begin{document}

\title{\ Style Transfer based Coronary Artery Segmentation in X-ray Angiogram}
\author{Supriti Mulay\textsuperscript{1,2}
\and
Keerthi Ram\textsuperscript{2}
\and
Balamurali Murugesan\textsuperscript{1,2}
\and
Mohanasankar Sivaprakasam\textsuperscript{1,2}\\
\textsuperscript{1} Indian Institute of Technology Madras, Chennai, India\\
\textsuperscript{2} Healthcare Technology Innovation Centre, Chennai, India\\
}


\maketitle
\ificcvfinal\thispagestyle{empty}\fi

\begin{abstract}
    X-ray coronary angiography (XCA) is a principal approach employed for identifying coronary disorders. Deep learning-based networks have recently shown tremendous promise in the diagnosis of coronary disorder from XCA scans. A deep learning-based edge adaptive instance normalization style transfer technique for segmenting the coronary arteries, is presented in this paper. The proposed technique combines adaptive instance normalization style transfer with the dense extreme inception network and convolution block attention module to get the best artery segmentation performance. We tested the proposed method on two publicly available XCA datasets, and achieved a segmentation accuracy of 0.9658 and Dice coefficient of 0.71. We believe that the proposed method shows that the prediction can be completed in the fastest time with training on the natural images, and can be reliably used to diagnose and detect coronary disorders.
    \end{abstract}

\section{Introduction}
Coronary disease is one of the leading causes of death worldwide, and X-ray coronary angiography (XCA) images are a \textit{gold standard} imaging procedure used by cardiologists to diagnose and treat coronary diseases. Since the XCA image is a projection of a 3D coronary artery structure on a 2D plane, the image is prone to an inherent artifact. Automatic segmentation of coronary arteries is an extremely useful technique in the diagnosis of coronary abnormalities. However, automatic segmentation of coronary artery vessels is challenging because of low contrast, high Poisson noise of low-dose X-ray imaging \cite{Xian}, and the artefacts resulting from XCA images projected in 2D. 

Deep learning (DL) approaches in medical imaging have recently demonstrated unprecedented progress with artificial intelligence \cite{Esteva}. Various DL based algorithms have recently outperformed the traditional image processing methods for object segmentation. Although DL achieves state-of-the-art segmentation performance \cite{yang}, the accuracy of these methods cannot be generalized. In other words, the segmentation accuracy of networks trained on a specific dataset does not extend to datasets of other modalities. The existing DL-based artery segmentation framework \cite{yang, Hao, Fernando, Samuel, simonyan} usually train an individual model for each dataset separately to segment the artery vessels. This approach is disadvantageous because these models need a lot of segmented coronary images for training, which are not easy to obtain. Also, the processing time required in many of the models \cite{Fernando, Hao} while applying the multiple filters is not practical. There is also a lot of variance in the shape and size of coronary arteries, and unless the models are trained using images that account for the diversity of angiography characteristics, these models do not generalize well. Therefore, it is highly desirable to train a model that does not need a lot of input images to train, can be easily generalized, and does not take a long time to run. To deal with the problems mentioned above, we present an algorithm that can segment the coronary artery vessels, and easily generalized. Using an Edge Adaptive Instance Normalization (Edge-AdaIN) style transfer network trained purely on natural images, we show that this network can be used to segment coronary arteries. 
The main contributions of the paper are: 
\begin{itemize}
  \item We designed \textit{edge detection} neural network techniques along with an \textit{attention module} in \textit{adaptive instance normalization} style transfer framework for artery vessel segmentations. The new Edge-AdaIN only needs few parameters and can be trained in a relatively short period of time.
  \item To the best of our knowledge, training on the natural images and testing with the medical images is a \textit{unique} and \textit{novel} approach for artery vessel segmentation. Vessel segmentation is thus possible without the prior knowledge of XCA images.
  \item This algorithm achieves promising real-time performance (20 fps) for 300x300 XCA images with promising accuracy. 
  
\end{itemize}

\noindent The rest of the paper is organized as follows. Section \ref{related_work} explores the related work for the proposed method. The proposed approach is described in Section \ref{method}. Further, Section \ref{experiment} summarizes the experimental setup and results. The ablation study experiment is described in Section \ref{ablation}. Finally, the discussion and conclusions are presented in Section \ref{discussion}.

\section{Related Work} \label{related_work}
\subsection{Coronary Artery Segmentation}
A deep learning model with a convolutional neural network is employed by Nasr-Esfahan \cite{Nasr}. Fernando \etal \cite{Fernando} constructed an autonomous segmentation of artery vessels, utilising the multiscale analysis, performed with Gaussian filters in the spatial domain and Gabor filters in the frequency domain. Jiang \etal \cite{Jiang} incorporated multiresolution and multiscale convolution filtering in U-Net network for artery segmentation. Two-vessel extraction layers were used in vessel-specific skip chain convolutional network by Samuel \etal \cite{Samuel}. U-Net based sequential vessel segmentation deep
network architecture called SVS-net is proposed by Hao \etal \cite{Hao} to segment artery vessels. AngioNet combines an angiographic processing network with a semantic segmentation network such as Deeplabv3+ to segment coronary arteries \cite{Iyer}.

However, the processing time required while applying the multiple filters was not practical, and it is not always possible to have a large number of medical images to train the network.

\subsection{Style Transfer}
Neural Style Transfer is the art of manipulating the image with the appearance style of another image. Adaptive instance normalization (AdaIN) \cite{Huang} style transfer is a state-of-the-art style transfer that simply aligns the channel-wise mean and variance of the content image to match those of the style image. Structure preserving style transfer network is employed by Cheng \etal 
with the addition of edge detection for local structure and depth prediction network for global structure refinement. A refine network is proposed by Zhu \etal \cite{Zhu} to preserve the details of content and overcome the distortion of stylized image. 

\section{Method}\label{method}
\begin{figure*}[!htb]
\begin{center}
   \includegraphics[width=0.8\linewidth]{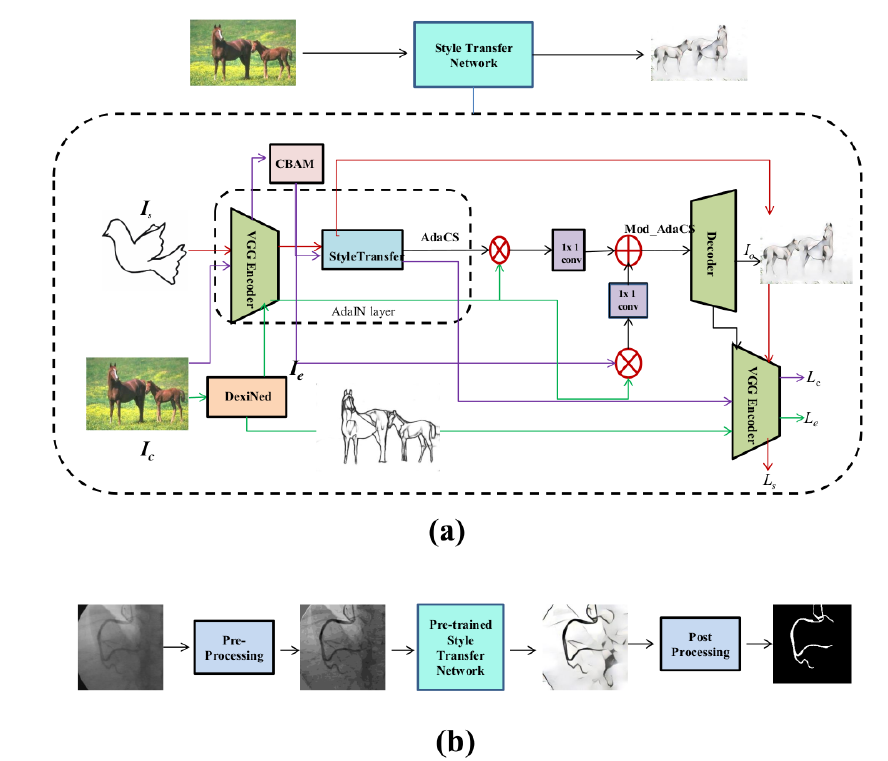}
\end{center}    
   \caption{Pipeline of coronary artery segmentation using Edge-AdaIN style transfer. (a) indicates training architecture with natural images, (b) demonstrates the testing with XCA images with pre-trained style transfer network.}
   {\label{fig:framework}}
\end{figure*}

Our method is influenced by adaptive instance normalization (AdaIN) \cite{Huang} layer as well as structure-preserving neural style transfer \cite{cheng}  framework, which are state-of-the-art style transfer networks. The proposed architecture is illustrated in Figure \ref{fig:framework}. Our coronary artery segmentation is divided into two sections viz: Pre-processing and Edge-AdaIN. The pre-processing step is explained in Section \ref{preprocessing}, and the Edge-AdaIN style transfer network step is explained in Section \ref{Edge_Ada}. 
\subsection{Pre-processing} 
\label{preprocessing}
Good spatial and contrast resolution in XCA images play an important role in coronary artery segmentation. Noise-free images are obtained using median filter and non-local means with statistical nearest neighbours algorithm \cite{Frosio}. The vessel structure intensity is enhanced by multiscale top-hat transform \cite{Hamid}. Multiscale top-hat transform acts as high- pass filter, extracts the bright areas of the image and removes the background of XCA image with enhancing the contrast. The key principle of multiscale top-hat transform is given as \cite{Hamid}
\begin{equation}
    I_{en} = I + I_{th}-I_{bh}
    \label{top_hat}
\end{equation}
where $I_{en}$ refers to the enhanced image, I is an original image, $I_{th}$ denotes the top-hat transform (bright area addition), and $I_{bh}$ denotes the bottom-hat transform (dark area subtraction). This equation has been adapted in the present work while enhancing the vessel structure.
\subsection{Edge-AdaIN style transfer}\label{Edge_Ada}
\subsubsection{AdaIN}
Style transfer is a technique that recomposes the content of an image into the style of another image. It is thus possible to get a boundary detection image with a specific style image. As aforementioned in Sec. \ref{related_work}, adaptive instance normalization (AdaIN)\cite{Huang} layer style transfer is a simple algorithm based on the encoder-decoder architecture. An encoder is fixed to the relu 4\textunderscore{1} layer of a pre-trained VGG-19 network\cite{simonyan}. The content and style images are encoded in feature space to get the feature map. The content feature maps are refined using a convolution block attention module. The content adaptive feature refinement maps and style feature maps are fed to AdaIN layer. The AdaIN layer aligns the channel-wise mean and variance of refined content image $c$ to match those of style image $s$, thus producing the target feature maps $AdaIN (c,s)$. The central equation of AdaIN is given as
\begin{equation}
    AdaIN(c,s) = \sigma(s)\: \left[\frac{c-\mu (c)}{\sigma (c)}\right] + \mu  (s)
    \label{AdaIN_eqn}
\end{equation}
\noindent where $c$ refers to the refined content image, $s$ denotes the style image, and $\sigma$ and $\mu$ are respectively the standard deviation and mean computed across the spatial dimensions independently for each channel. We embrace AdaIN layer as a core component of our Edge-AdaIN network.
\subsubsection{Convolution Block Attention Module (CBAM)}
\begin{figure}[!htbp]
\begin{center}
   \includegraphics[width=1.1\linewidth]{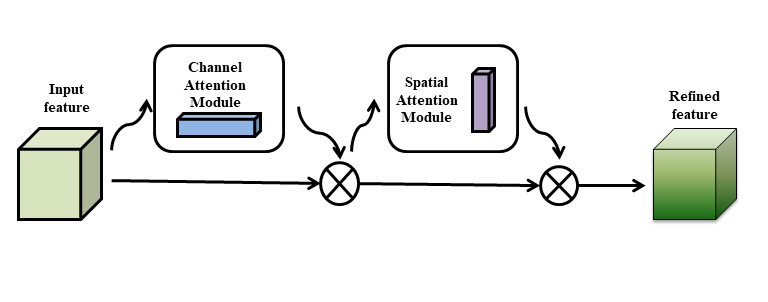}
\end{center}
   \caption{Convolution block attention module structure}
{\label{fig:cbam}}
\end{figure}

Convolution block attention module is proposed by Woo \etal \cite{cbam} to boost the feature maps. Figure \ref{fig:cbam} explains the CBAM structure, consisting of two sequential sub-modules, viz., channel attention and spatial attention. The channel attention essentially provides a weight for each channel, making up the filters to learn the small values and conveys the important feature map for learning. The spatial attention generates a mask enhancing those features that define the structure and fetch what is essential to learn within the feature map. The combination of these sub-modules thus enhances the feature maps by improving the performance.\\

\noindent The focus for object detection should be more on the features than the background information. We thus adopt CBAM for encoded content feature maps for their enhancement purpose. 

\subsubsection{Structure preserving network}
The local structure enhancement is a crucial aspect of coronary artery segmentation. We thus introduce an edge detection network as a local structure refinement network. We take the \textit{dense extreme inception network} (DexiNed) (\cite{DexiNed}) approach as the edge-preserving network, which is an end-to-end approach generating a thin edge-map as output. It is computationally inexpensive and extracts more edge feature information with an inception network backbone, thus enabling DexiNed to predict highly accurate and thin edge images. This approach can be used in any edge detection task without any previous training or a fine-tuning process. It thus efficiently generate edge maps with multi-level perceptual features demonstrating promising results for XCA images though trained on BIPED dataset\footnote{https://www.kaggle.com/xavysp/biped}.\\ 
In our implementation, we use its edge map response as local structure I\textsubscript{e}.

\noindent Our Edge-AdaIN style transfer network \textit{T} takes refined content feature map $\mathcal{C}$, encoded edge map $\mathcal{E}$, and encoded style feature map $\mathcal{S}$ as inputs, and integrates an output image \textit{Mod\textunderscore{AdaCS}} as given in Eq. (\ref{mod_ada_eqn}), recombining the content, edge of the content, and the style image.
\begin{equation}
    Mod\_{AdaCS} = f(f(\mathcal{C}),f(\mathcal{S}),f(\mathcal{E}))
    \label{mod_ada_eqn}
\end{equation}
A randomly initialized decoder $g$ is prepared to plan \textit{T} back to the image space, thus creating the stylized image \textit{T}($\mathcal{C}$,$\mathcal{S}$,$\mathcal{E}$) as
\begin{equation}
    T(\mathcal{C},\mathcal{S},\mathcal{E}) = g(Mod\_{AdaCS})
    \label{decode_eqn}
\end{equation}
where the decoder is a mirror image of the encoder. We introduced a minor change in the encoder-decoder activation function as LeakyReLU instead of ReLU in the original AdaIN layer \cite{Huang}.

\noindent Stylized image \textit{T}($\mathcal{C}$, $\mathcal{S}$, $\mathcal{E}$), with a specific boundary style, is a boundary detected image with some artifacts. An extra step, the \textit{morphological} post-processing, is thus required to transform the stylized output into a binary mask achieving the desired result. 

\section{Experiments and Results}\label{experiment}
\subsection{Implementation Details}
We train our network with BSDS500 dataset \cite{BSDS} as content images, and some images of WikiArt painting and boundary sketch drawing as style images. The content image dataset contains BSDS 200 training images. Adam optimizer ( \cite{Adam}) is used with a batch size of 1. We randomly crop regions of size 256×256 during training. Since our network is fully convolutional, it can be applied to images of any size during testing. All the experiments were carried out on NVIDIA GTX 1080 8 GB GPU on a system with 16GB RAM and Intel Core-i5 7th generation @3.20GHz processor, and the network was implemented on Pytorch. Availability of extensive medical data is not always possible. We thus trained the network with natural images and not with X-ray images for style transfer. 

\noindent We use two databases \cite{Fernando} of 134 XCA images and \cite{Hao} of 30 XCA images for testing and corresponding ground truth images. The test image dimensions are 300x300 pixels and 512x512 pixels, respectively. We used the pre-trained VGG-19 \cite{simonyan}, similar to AdaIN (\cite{Huang}), to compute the loss function while training the decoder as
\begin{equation}
    \mathcal{L} = \alpha.\mathcal{L}_{c} + \beta.\mathcal{L}_{s} + \gamma.\mathcal{L}_{e}
    \label{total_loss_eqn}
\end{equation}
which is a weighted combination of the content loss $\mathcal{L\textsubscript{c}}$, style loss $\mathcal{L\textsubscript{s}}$, and the edge loss $\mathcal{L\textsubscript{e}}$ with weights $\alpha$, $\beta,$ and $\gamma$, respectively. The content loss is Euclidean distance between the target features and the features of the output image. We use AdaIN layer output AdaCS as the content target instead of the content image feature similar to existing AdaIN framework. The content loss $\mathcal{L\textsubscript{c}}$ is computed as
\begin{equation}
    \mathcal{L}_{c}=\left \|\: f[g(Mod\_AdaCS)]\:-\: AdaCS \:\right \|_{2}
    \label{content_loss_eqn}
\end{equation}
\noindent The style loss is calculated between the mean and standard deviation of the style features and target features as
\begin{equation}
\left.
\ba{lll}
\displaystyle \mathcal{L\textsubscript{s}} = \sum_{i}^{L}\left \|\: \mu \left\{\phi _{i}[g(Mod\_AdaCS)]\right\}-\mu_{i} [\phi_{i}(s)]\: \right \|_{2} + \\[2ex]
    \sum_{i}^{L}\left \|\: \sigma \left\{\phi_{i} [g(Mod\_AdaCS)]\right\}-\sigma [\phi_{i} (s)] \:\right \|_{2}
\ea
\right\}
\label{style_loss_eqn}
\end{equation}
where each $\phi_i$ denotes a layer in VGG-19 used to compute the style loss. Four layers of the decoder, with equal weights, are used in the style loss computation. 

\noindent The edge loss is similarly computed as a sum of the absolute difference between the target features and edge features given as
\setlength{\abovedisplayskip}{3pt}
\setlength{\belowdisplayskip}{3pt}
\begin{equation}
    \mathcal{L\textsubscript{e}} = \sum_{i}^{L}\left \|\: \left\{\phi _{i}[g(Mod\_AdaCS)]\right\} - \left\{\phi_{i}[f(e)]\right\} \:\right \|_{2} 
\label{edge_loss_eqn}
\end{equation}
where $f(e)$ is an encoded edge feature map of DexiNed edge output. The edge loss, similar to style loss, is computed for all four layers of VGG-19. The weighting parameters, from Eq. (\ref{total_loss_eqn}), used in our experiments are $\alpha = 1,$ $\beta = 0.05,$ and $\gamma = 0.05$ .

\subsection{Results}

\subsubsection{Combination of CBAM and DexiNed module analysis}
\begin{figure}[H]
\begin{center}
   \includegraphics[width=\columnwidth]{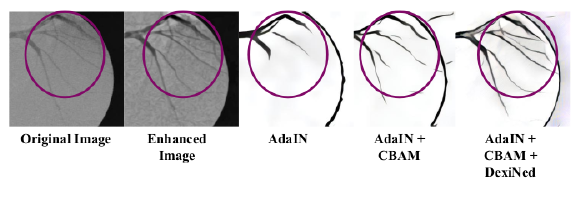}
\end{center}
   \caption{Experiment with CBAM and DexiNed module with AdaIN. From left to right (Original image, Enhanced image, style transferred output with AdaIN \cite{Huang}, with combination AdaIN and CBAM module, with the combination of AdaIN, CBAM, and DexiNed (proposed) network). The proposed method has recovered the structure (pink circle), but the existing module and module with CBAM only partially detect the structure.}
{\label{fig:CBAM_exp}}
\end{figure}

It is indeed interesting to demonstrate, as shown in Figure \ref{fig:CBAM_exp}, the enhanced local structure refinement from the content image combining CBAM framework (\cite{cbam}) with an edge detection DexiNed (\cite{DexiNed}) network. Qualitatively, the result obtained by the combination of AdaIN, CBAM, and DexiNed is significantly better than the one obtained by AdaIN alone (as illustrated in Figure \ref{fig:CBAM_exp}). The proposed method keeps the structural consistency because of the edge map and attention module.

\subsubsection{Comparison with other methods}

The presented approach in this manuscript is compared with the three other coronary artery segmentation methods from the literature: 1) Multiscale multilayer perceptron (MLP) based method (\cite{Fernando}), 2) Vessel specific skip chain convolutional(VSSC) Net based methods \cite{Samuel}, and 3) Sequential  vessel segmentation via deep channel attention network (SVS-Net) \cite{Hao}. We have used the two test dataset of 30 images from the database given in \cite{Fernando} used in the first two methods, and  30 test images given in \cite{Hao} used in the third method. We used the trained network file of the BSDS500 dataset with natural images for style transfer. \\

\noindent \textbf{\underline{Qualitative Analysis}}:\\

\begin{figure}[!htbp]
\begin{center}
   \includegraphics[scale=0.7]{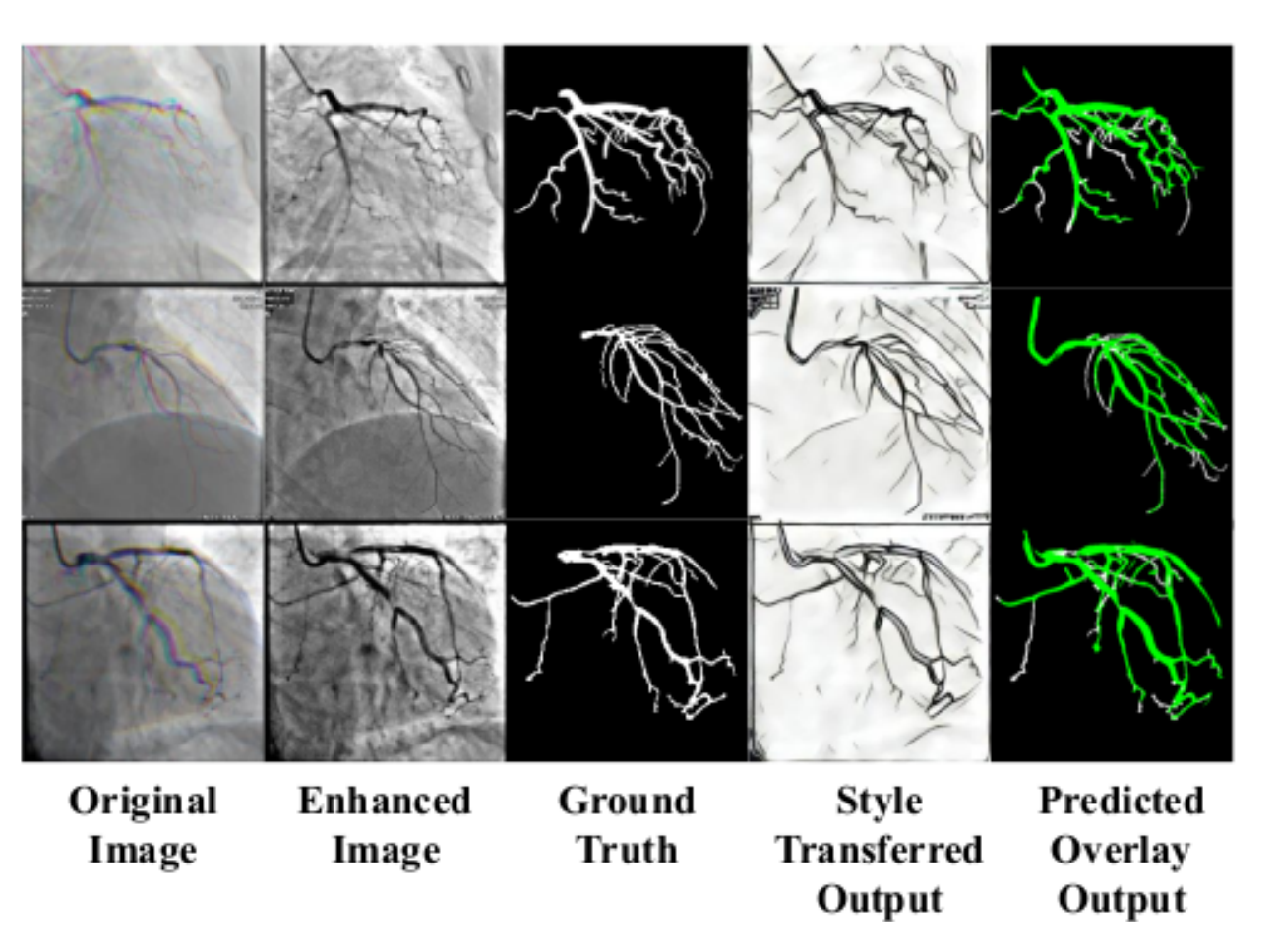}
\end{center}
   \caption{Qualitative analysis of Edge-AdaIN with 512x512 XCA database \cite{Hao}. The last column shows the overlaid predicted output in green on the ground truth map in white color.}
{\label{fig:svs}}
\end{figure}

\noindent We observed for XCA images \cite{Hao}, of 512×512 resolution with 8 bits per pixel, our method prediction is comparable with ground truth. Figure \ref{fig:svs} shows the comparison of our method prediction with ground truth. Though vessels with plenty of branches and poor visibility are there, Edge-AdaIN has still detected most branches. 

\begin{figure*}[t!]
\begin{center}
   \includegraphics[scale=2.7]{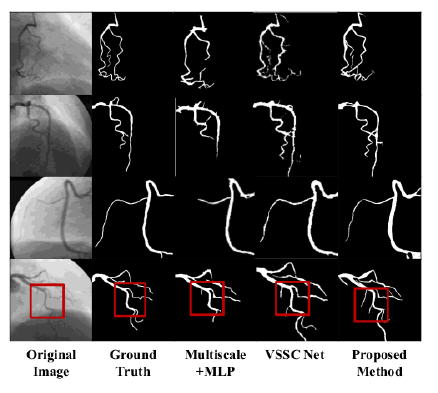}
\end{center}
   \caption{Qualitative analysis of Edge-AdaIN compared to other deep learning methods with 300x300 XCA database \cite{Fernando}. We can see that the proposed method segmentation output is close to the ground truth images. The last row of red boxes indicates vessel structure not marked in ground truth is detected in VSSC-Net and our proposed method.}
{\label{fig:dca}}
\end{figure*}

\noindent Figure \ref{fig:dca} demonstrates the comparison of the segmented result obtained by the various methods. All the predicted artery segmentation results are obtained with our method after post-processing of style transfer output. It is important to note that, we have not used any X-ray images during the training of style transfer network, while the training is performed on the same X-ray database (\cite{Fernando}) images of 300x300 resolution in the first two methods. Similarly, the same training and testing XCA database(\cite{Hao}) of 512x512 resolution images is used in the third method. Nevertheless, the predicted output by our method is closely matching with that of the other methods. In some cases (e.g. row 4 in Figure \ref{fig:dca}), we observed that our method had detected the arteries present in the original X-ray image but not in the ground truth and multiscale MLP based method. This behaviour is observed because the edge detection with DexiNed (\cite{DexiNed}) in the proposed method detect all edges in the images. 
\\
\\
\noindent \textbf{\underline{Quantitative Analysis}}:\\
\noindent We quantitatively evaluated our Edge-AdaIN method with the other deep learning methods in this section. The evaluation is performed on the coronary artery segments by the predicted results and the ground truth label with evaluation metrics, namely, Accuracy, Sensitivity, Specificity, and Dice Coefficient defined as
\begin{equation}
\left.
\begin{array}{lll}
\displaystyle Accuracy = \frac{(tp +tn)}{(tp+fp+tn+fn)}, \\[4ex] \displaystyle Sensitivity = \frac{tp}{(tp + fn)}, \:Specificity = \frac{tn}{(fp+tn)}, \\[4ex]
\displaystyle Dice = \frac{(2 * Precision * Sensitivity)}{(Precision + Sensitivity) },\\[4ex]
\displaystyle \mbox{where}\:\: Precision = \frac{tp}{(tp+fp)}
\end{array}
\right\}
\label{metric_eqn}
\end{equation}
where $tp$ = number of correct foreground pixels, $fp$ = number of incorrect foreground pixels, $tn$ = number of correct background pixels, and $fn$ = number of incorrect background pixels.\\

\noindent Table \ref{tab:comparison} demonstrates the comparison between the segmentation results of Edge-AdaIN and other existing methods on the database \cite{Fernando}. It can be seen from Table \ref{tab:comparison} that the performance of our method is at par with the other existing methods though we have used natural images for training and not XCA images. The metric values are thus fairly comparable with the other deep learning methods trained on XCA images. 
\begin{table}[htbp]
  \centering
  \caption{Comparison of segmentation results of our Edge-AdaIN method and other methods on the 30 XCA \cite{Fernando} images.}
    \vspace*{3mm}
    \resizebox{\columnwidth}{!}{%
    \begin{tabular}{p{2.215em}ccccc}
    \toprule
    \multicolumn{2}{p{5em}}{\textbf{Method}} & \textbf{Sensitivity} & \textbf{Specificity} & \textbf{Accuracy} & \textbf{Dice} \\
    \midrule
    \midrule
    \multicolumn{2}{p{5em}}{Multiscale+MLP \newline{} \cite{Fernando}} & 0.6364 & 0.988 & 0.9698 & 0.6857 \\
    \multicolumn{2}{p{5em}}{VSSC Net \cite{Samuel}} & 0.7634 & 0.9857 & 0.9749 & 0.7738 \\
    \multicolumn{2}{p{5em}}{Single U-Net \cite{Ronneberger}} & 0.7165 & 0.9815 & 0.9645 & 0.7571 \\
    \multicolumn{2}{p{5em}}{Multiresolution U-Net \cite{Jiang}} & 0.7978 & 0.9885 & 0.9765 & 0.7905 \\
    \multicolumn{2}{p{5em}}{Edge-AdaIN(30 image test)} & 0.7867 & 0.9756 & 0.9658 & 0.7165 \\
    \bottomrule
    \end{tabular}%
    }
  \label{tab:comparison}%
\end{table}%

\begin{table}[htbp]
  \centering
  \caption{Comparison of Edge-AdaIN with SVS-Net database \cite{Hao}}
   \vspace*{3mm}
    \resizebox{\columnwidth}{!}{
    \begin{tabular}{p{7.145em}p{4.645em}p{4.785em}p{4.785em}}
    \toprule
    \textbf{Segmentation Method } & \textbf{Detection Rate } & \textbf{Precision} & \textbf{F-measure} \\
    \midrule
    \midrule
    \multicolumn{1}{c}{SVS-Net 2D + CAB} & 0.7638± 0.0738 & 0.8595± 0.0684 & 0.8046± 0.0459 \\
    Edge -AdaIN & \textbf{0.7906± 0.1107} & 0.6588± 0.1204 & 0.7146± 0.0747 \\
    \bottomrule
    \end{tabular}%
    }
  \label{tab:svs_metric}%
\end{table}%

\noindent Similarly, we compared our method with SVS-Net on the database \cite{Hao}. This database includes extremely low-contrast vessels, and vessel trees contain a lot of thin vessel branches. Table \ref{tab:svs_metric} exhibit the comparison of our approach and the SVS-Net method. It can be perceived from Table \ref{tab:svs_metric} that our approach has a better detection rate as compared to SVS-Net. The proposed method's performance for precision and F-measure  is moderately close to the existing method despite the complexity of XCA images. \\

\noindent \textbf{\underline{Speed Analysis}}:
The speed of our method is compared with the other artery segmentation methods in Table  \ref{tab:exec_time}. Most of our processing time is spent on content encoding, style encoding, decoding, and edge detection. Our algorithm runs on 300 x 300 XCA images with an average execution time of 0.05 seconds lowest execution time of all the other segmentation algorithms. It is evident that, the filter-based and patch-based algorithms are computationally expensive compared with our proposed approach. This speed can be further improved by employing more efficient architecture. For 512x512 images also our method shows 0.1 seconds execution time which 10 fps, which is good for real-time execution. 

\begin{table}[htbp]
  \centering
  \caption{Average execution time for segmentation method for each image}
    \resizebox{\columnwidth}{!}{%
    \begin{tabular}{p{4.215em}cccc}
    \toprule
    \multicolumn{3}{c}{\textbf{Segmentation Method}} & \multicolumn{2}{p{9.285em}}{\textbf{Execution Time/image(s)}} \\
    \midrule
    \midrule
    \multicolumn{3}{p{10.215em}}{Multiscale+MLP \cite{Fernando}} & \multicolumn{2}{c}{1.89} \\
    \multicolumn{3}{p{10.215em}}{VSSC Net \cite{Samuel}} & \multicolumn{2}{c}{0.1} \\
    \multicolumn{3}{p{10.215em}}{Single U-Net \cite{Ronneberger}} & \multicolumn{2}{c}{0.19} \\
    \multicolumn{3}{p{10.215em}}{Multiresolution U-Net \cite{Jiang}} & \multicolumn{2}{c}{0.39} \\
    \multicolumn{3}{p{10.215em}}{Edge-AdaIN} & \multicolumn{2}{c}{\textbf{0.05}} \\
    \multicolumn{3}{p{18.575em}}{SVS-Net \cite{Hao} ( 512x 512 pixels)} & \multicolumn{2}{c}{0.178} \\
    \multicolumn{3}{p{18.575em}}{Edge-AdaIN (512x512 pixels)} & \multicolumn{2}{c}{\textbf{0.1}} \\
    \bottomrule
    \end{tabular}%
  \label{tab:exec_time}%
  }
\end{table}%

\section{Ablation Study} \label{ablation}

An ablation study experiment is designed better to understand the style transfer mechanism for XCA images. 
\subsection{Comparison with other style transfer methods}
\begin{figure*}[htbp!]
\begin{center}
  {\includegraphics[scale=3.0]{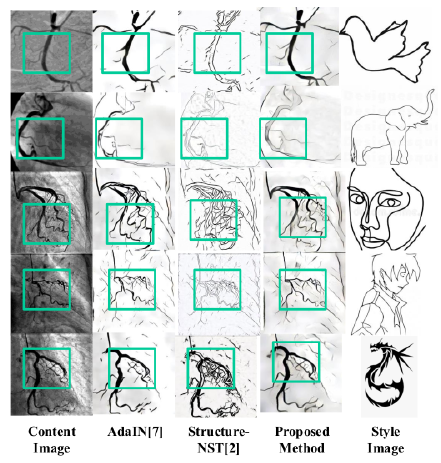}}
\end{center}
   \caption{Qualitative analysis of proposed Edge-AdaIN in comparison to other style transfer methods. The structures in the region given by green box has been sharply detected by Edge-AdaIN; AdaIN, Structure-NST has detected the structures but not with correct thickness and continuity.}
{\label{fig:style_comp}}
\end{figure*}
We evaluate the effectiveness of the proposed Edge-AdaIN style transfer method with the state-of-the-art style transfer methods existing in the literature. We chose a style image such that the content image boundaries get transferred to a boundary image by feature statistics. Figure \ref{fig:style_comp} illustrate the stylized images with our method and the other state-of-the art style transfer methods. AdaIN (\cite{Huang}) method adjusts the mean and variance of the content features to stylized images, but the detailed structure is not preserved. The structural consistency between the original images and transferred images is maintained by the structure-preserving neural style transfer method \cite{cheng}, where the image representation displays a skeleton-like structure for arteries. The proposed method, in contrast, changes the boundary style patterns while maintaining the local structure useful for segmenting the coronary artery.\\

\subsection{Style image selection}
\begin{figure}[H]
\begin{center}
   \includegraphics[scale=2]{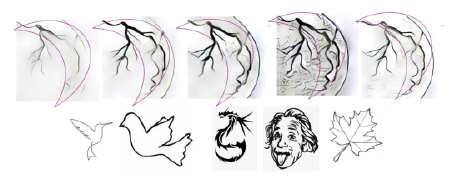}
\end{center}
   \caption{Difference in style transferred images depending on the style image. Structure given in pink dotted lines shows the difference in line thickness and continuity with style image choice.}
{\label{fig:style_comp_same_img}}
\end{figure}

In the style transfer process, the content and reference style image are blended. The resulting output image retains the core element of the content image along with the style of a reference image. A stylized image will give the boundary structure of the input image, if the style image has a boundary outline like structure. Different style images are selected in Figure \ref{fig:style_comp_same_img}, having varying thicknesses. A larger thickness style image (bird 2) shows better results with the proposed method for artery or vessel-like structures for the dataset \cite{Fernando}. The results are the same with a small thickness style image (first bird), but the small vessel thickness will differ with segmented output. Structure continuity is also varying based on the style image chosen. More background structure is getting changed with Einstein image. Thus depending on content image structure style image should be chosen. 

\noindent We have chosen two different style transfer images( bird and girl face image) for two XCA databases \cite{Fernando} and \cite{Hao}, as the thickness of arteries and vessel tree structure is differing in these databases. For vessel tree branches dataset \cite{Hao}, girl face image demonstrated superior results as compared to other boundary style images. It is thus seen that, choosing the relevant style image is important for the proposed Edge-AdaIn algorithm.  In the structure-preserving neural style transfer method, we need to train the network every time the style image changes. However, in our case, though the network is not trained with the particular style image, we can still get the results with unrevealed style images.
\section{Discussion and Conclusion}
\label{discussion}
A novel style transfer based artery segmentation method trained on natural images is proposed in this paper. We changed the content of an image in one domain to the style of an image in another domain (boundary line structure) to segment the vessel structure from XCA images. We showed that the coronary arteries can be segmented in a time frame (20 fps or 0.05 seconds) compatible with surgical procedures performed using XCA images. Although we trained on the natural images, coronary artery segmentation on X-ray images is still comparable with other deep learning methods. 

\noindent We also studied the effect of different boundary patterns and observed that the choice of style images affected the segmentation accuracy. Therefore one can conclude that by changing the style images, this algorithm can be easily extended to segment images for various applications. In future work, we plan to explore the applicability of the proposed algorithm for retinal vessel extraction. We also plan to examine Edge-AdaIN style transfer on other medical image segmentation problems.

\noindent While further work is needed to validate our approach in a clinical setting, it is our belief that our approach can save considerable time in diagnosis, detection and treatment of cardiac diseases.
\newline
\noindent\textbf{Acknowledgement}
We would like to thank Prof. Shantanu Mulay  and Dr. Prashanth Dumpuri for providing their valuable comments and suggestions. 

{\small
\bibliographystyle{ieee_fullname}
\bibliography{Style_transfer_artery_seg.bib}
}

\end{document}